\documentclass[letterpaper,twocolumn,10pt]{article}
\usepackage{usenix-2020-09}

\usepackage[frozencache,cachedir=_minted-output]{minted}
\usepackage{xcolor}
\usepackage{listings}
\usepackage{graphicx}
\usepackage{subcaption}
 
\usepackage{mathptmx}
\usepackage{amsmath}
\usepackage{enumitem}
\usepackage{multirow}
\usepackage{subcaption}
\usepackage{url}

\usepackage{comment}
\usepackage{pifont}

\usepackage{color}
\usepackage{xspace}
\usepackage{booktabs,tabularx}
\usepackage{ragged2e}
\usepackage{tikz}
\usepackage{dcolumn}
\usepackage{dsfont}
\usepackage{graphicx}
\usepackage{float}

\usepackage{listings}
\usepackage{tcolorbox}
\usepackage{xcolor}
\usepackage{enumitem}
\usepackage{etoolbox}

\usepackage{hyperref}
\usepackage{flushend}
\usepackage{siunitx}

\usepackage{wasysym}

\newif\ifshowtodos
\showtodosfalse %

\usetikzlibrary{positioning}
\usetikzlibrary{decorations.pathreplacing}
\newcommand\code[1]{\texttt{#1}}
\newcommand{\toolname}{\textit{SplittingSecrets}}
\newcommand{\DMP}{DMP}

\newif\iftrim
\trimfalse %

\begin{document}
\date{}

\title{\Large \bf \toolname{}: A Compiler-Based Defense for Preventing\\ Data Memory-Dependent Prefetcher Side-Channels}

\author{
  {\rm Reshabh K Sharma}\\
  University of Washington
\and
    {\rm Dan Grossman}\\
    University of Washington
\and
    {\rm David Kohlbrenner}\\
    University of Washington
    }

\maketitle

\begin{abstract}

Traditional side-channels take advantage of secrets being used as inputs to unsafe instructions, used for memory accesses, or used in control flow decisions.
Constant-time programming, which restricts such code patterns, has been widely adopted as a defense against these vulnerabilities.
However, new hardware optimizations in the form of \emph{Data Memory-dependent Prefetchers} (DMP) present in Apple, Intel, and ARM CPUs have shown such defenses are not sufficient.
These prefetchers, unlike classical prefetchers, use the content of memory as well as the trace of prior accesses to determine prefetch targets. An adversary abusing such a prefetcher has been shown to be able to mount attacks leaking \emph{data-at-rest}; data that is never used by the program, even speculatively, in an unsafe manner.

In response, this paper introduces \toolname{}, a compiler-based tool that can harden software libraries against side-channels arising from DMPs. \toolname{}'s approach avoids reasoning about the complex internals of different DMPs and instead relies on one key aspect of all DMPs: leakage requires data to resemble addresses.
To prevent secret data from leaking, \toolname{} transforms memory operations to ensure that secrets are never stored in memory in a manner resembling an address, thereby avoiding DMP activation on those secrets.
Rather than disable a DMP entirely, \toolname{} can provide targeted hardening for only specific secrets entirely in software.

We have implemented \toolname{} using LLVM, supporting both source-level memory operations and those generated by the compiler backend for the AArch64 architecture,
We have analyzed the performance overhead involved in safeguarding secrets from DMP-induced attacks using common primitives in libsodium, a popular cryptographic library when built for Apple M-series CPUs.
\end{abstract}

\section{Introduction}

The pursuit of performance in general-purpose CPUs has lead to clever and surprising optimizations.
Many of these optimizations have significant trade-offs, particularly concerning unintended information leakage through microarchitectural side-channels~\cite{uarch-side-channel, uarch-side-channel-1} varying from caches~\cite{cache-sc, cache-sc-1} to TLBs~\cite{tlb-sc, tlb_bleed}, branch predictors~\cite{branch-sc, branch-sc-1}, on-chip interconnects, memory management units, speculation~\cite{spec-sc-1, spectre}, and more.

Most microarchitectural side-channels leak data \emph{in-use}~\cite{pandora} --- data that is either speculatively~\cite{spectre, meltdown} or non-speculatively~\cite{cache-sc, cache-sc-1} passed to unsafe instructions.
Such vulnerabilities leak secrets when they influence which addresses are in cache~\cite{prime_probe, flush_flush, flush_reload} and/or when secrets are used as inputs to instructions that vary execution time based on operands~\cite{grossschadl2010multiply,FPU_leaky}.
Constant-time (CT) programming attempts to prevent such leaks by ensuring that secrets are never used in instructions known to leak a function of their operands.

Unlike data in-use side-channels, data \emph{at-rest} side-channels occur when a program sequence leaks data that is not computed on, even speculatively.
The most notable example of this is recent work using data memory-dependent prefetchers (DMPs)~\cite{augury, gofetch, peek-a-walk-wang-2025}.
These are prefetchers that depend not only on previous memory accesses the program makes, but also on contents of memory~\cite{dmp-1, dmp-2, dmp-3} to fetch additional data.
Such prefetchers are known to be deployed in the Apple A- and M-series processors and Intel 13th+ generation CPUs.
We discuss the design of data memory-dependent prefetchers, the specific families we consider for this work, and how they have been shown to allow leaking secrets in Section~\ref{sect:bkg}.

Our proposed defense uses a key requirement of any DMP side-channel: the prefetch must result in an at least partially successful page walk~\cite{peek-a-walk-wang-2025}.
If all secrets in memory are prefixed with a value that ensures no page walk succeeds (or even that no prefetch is started) no secrets will be leaked.
This can occur because either the prefetcher has a heuristic check for what qualifies as a likely address, or because the prefix is an unmapped (or unmappable) memory region.

To accomplish this, we rewrite how secrets are stored in memory.
Each secret is split into 32-bit chunks and each chunk is placed in the low half of a 64-bit word whose high half is a chosen prefix that does not satisfy the requirements for the DMP being considered.
As a result every aligned value in memory containing secret data fails address translation or is never even attempted to be prefetched.
This necessitates rewriting loads as well, as they need to reconstruct split values.
Our approach avoids modeling intricate details of DMPs, and is trivially adaptable to all known deployed DMPs simply by changing the chosen prefix.
We believe that future DMP implementations will follow a similar structure, and thus our defense will also be applicable to them.

While vendors have introduced some hardware configuration mechanisms (ARM Data Independent Timing --- DIT, and Intel Data Operand Independent Timing --- DOIT) that can disable DMP behavior on some processors, none of these mechanisms are consistently available in userspace or consistently disable DMP activation across all processors.
In line with prior work~\cite{cio} we provide sensitive software with a software-only defense that functions across all known DMP-enabled CPUs and with no privilege requirements.
Further, given that DMPs exist to provide performance improvements, it is desirable to have a targeted solution that only protects cryptographic secrets without completely disabling the optimization across the CPU and without dependence on the vendor's interpretation of the severity of the threat.

\toolname{}, our compiler-based tool, automatically performs this memory transformation approach, shielding secrets against side channels induced by DMPs.
Our method operates exclusively at the memory operation level and does not require changes to other operations.

Our key contributions are:
\begin{itemize}
    \item Proposal and design of the first compiler-based (and multi-platform) mitigation strategy for \emph{address-based} DMP data-at-rest side-channels.
    \item Implementation of our approach within LLVM~\cite{llvm} for ARM64, targeting cryptographic libraries.
    \item Analysis of the performance and memory overheads introduced by our memory representation transformation using the libsodium~\cite{libsodium} cryptographic library.
\end{itemize}

Our implementation along with tests and microbenchmarks used in analysis are  available at \href{https://github.com/orgs/dmp-mitigation}{https://github.com/orgs/dmp-mitigation}.

\section{Background}
\label{sect:bkg}

\subsection{Classical Prefetchers}

Classical prefetchers~\cite{classical-prefetcher} are designed to improve the efficiency of CPU memory accesses by predicting future data needs and fetching the data into cache before it's directly requested by the CPU~\cite{data-locality}.

Significant prior work has shown that classical prefetchers can also expose vulnerabilities.
Generally, these exploit timing differences in future cache accesses induced by prefetchers, where prefetcher actions are caused by secret-dependent behavior.
For example, stride prefetchers have been shown to be useful for amplifying cache attacks on the same core~\cite{prefetcher_attacks_svf,prefetcher_attacks_unveiling} or, less commonly, cross-core~\cite{prefetcher_attacks_prefetchx}.
Classical prefetchers can also serve as covert-channel mechanisms~\cite{prefetcher_attacks_a_fetching_tale, prefetcher_attacks_afterimage}.

A critical element of \emph{all} classical prefetchers is they operate only on the trace of addresses accessed by the program.
Thus, classical prefetcher attacks only work on code patterns that violate existing constant time programming rules.

\subsection{Data Memory-dependent Prefetchers}

A data memory-dependent prefetcher (DMP) uses both the prior memory access pattern \emph{as well as data returned from those accesses} to make more informed guesses about future memory requests~\cite{dmp_gretch, dmp_informed_prefetching, dmp_stateless_content_directed, dmp_tempo, dmp-1, dmp-2, dmp-3,huberty2018content}.

There are two types of data such prefetchers can operate on: addresses and indices.
An index-based prefetcher expects memory will contain array indices or offsets, and that the prefetcher will need to infer what the base-address is as well as compute the final likely address target.
Address-based prefetchers are simpler, and expect memory will contain virtual addresses that can be directly prefetched without additional computation.

All known examples of deployed DMPs in CPUs~\cite{augury,gofetch,peek-a-walk-wang-2025} are address-based prefetchers, and we use ``DMP'' to refer to an address-based DMP unless noted.

\subsubsection{The Security Implications of DMPs}

Unlike classic prefetchers, because DMPs use values in memory to determine what to prefetch, they allow new attacks against software written to current constant-time principles.

DMPs can be exploited when they choose to prefetch a target address derived from a secret, e.g. a cryptographic key value. Once prefetched, attackers can use standard cache side-channel techniques, such as PRIME+PROBE~\cite{prime_probe}, to determine which cachelines were prefetched and learn (partially) the value used for the prefetch address.

This differs from classical prefetcher behavior, which will not make a prefetch decision based on a secret unless program flow already makes a memory access based on a secret.

Recent work~\cite{augury,gofetch,peek-a-walk-wang-2025} has demonstrated this behavior can recover keys from implementations of RSA, DH key-exchange, post-quantum ML-KEM, as well as attack ASLR and recover kernel memory.
The cryptographic attacks assume a local adversary able to measure cache state via standard side-channels, and able to send chosen ciphertexts to the target program.
Ciphertexts are chosen to cause a memory operation the victim program executes to store a value $X$ or $Y$ \emph{depending on a specific bit of the key}.
When constructed correctly, $X$ passes the DMP validity check, and results in a prefetch of virtual address $X$, while $Y$ does not pass such a check.
Critically, this intermediate state is not semantically a pointer, not used to derive loaded memory addresses, and would not cause information leakage without a DMP.
This work addresses these attacks.

\subsubsection{Defending Against DMPs}

Vendors have tried to improve security against microarchitectural side-channel attacks broadly by providing configuration bits that guarantee (a subset of) instruction timings are operand-independent.

Since ARMv8a, ARM includes a Data Independent Timing (DIT)~\cite{ARMDIT} standard where any exception level can activate DIT by setting the PSTATE.DIT configuration bit during runtime.
DIT does not make clear if behavior such as the DMP should be covered or not.
It is also possible to disable Apple's DMP via undocumented configuration bits~\cite{Asahi,gofetchweb}, which Asahi Linux supports for disabling the DMP globally.
 
Intel's  Data Operand Independent Timing (DOIT) standard~\cite{intel_doit} (2022) is similar to ARMs DIT.
Unlike DIT, DOIT can only be enabled by the kernel but explicitly affects DMPs (Data-Dependent Prefetchers~\cite{intel_ddp} in Intel parlance.)

These mechanisms are inconsistent and ambiguous and not an effective mitigation plan for software that has minimal control over its execution environment.
ARM's DIT affects Apple M3 but not M1/M2~\cite{augury,gofetch}.
No OS we are aware of has support for userspace to request DOIT to be enabled.

Two prior works have sketched out, but not developed or evaluated, a defense with the same intuition as our system.
Augury~\cite{augury} notes that storing secrets only in the lower bits of each 64-bit aligned region should protect against the Apple DMP.
Similarly, ZebraFix~\cite{zebrafix} proposed a modification of their defense for mitigating DMP side-channels that uses data interleaving.

\subsubsection{DMPs in the Wild}

All known deployed DMPs operate on addresses, not indices, but vary in if they track and respond to prior prefetch accuracy, what history they maintain, and heuristics they apply.

Our mitigation strategy is deliberately constrained to stable, cross-generational characteristics, specifically inspecting aligned 64-bit words and the necessity of partial page walks. If future designs accept more high-bit patterns we can rotate prefixes, or reduce the secret fragment size (e.g., to a lower 16-bit slice) so reconstructed words remain implausible addresses.

\section{Threat Model}

Our work aims to harden software libraries that were safe to run on a system without a DMP, but are not safe on a system with a DMP.
We specifically target cryptographic software assumed to use constant time programming techniques and implement appropriate speculative attack defenses but is not aware of the threat posed by DMPs.

We assume our target program is running on a 64-bit system with an address-based DMP, similar to either the Apple M-series DMP or the Intel 13th gen DDP.
Both of these DMPs examine memory to identify likely pointers and dereference them, though they take different approaches as to when and how they perform this analysis.
To encompass all types of address prefetchers, we consider an aggressive implementation of the DMP where any piece of memory can be examined by the DMP at any time, and the target value being examined need not be a valid virtual address.
We assume that, as an over-approximation, the DMP only leaks when the value considered results in a partially successful page walk.

Our adversary can use both a classical cache side-channel with or without shared memory\cite{prime_probe, flush_reload}, as well as a page-walk side-channel~\cite{peek-a-walk-wang-2025,mmu_cache_attacks}. This implies that, for our threat model, \emph{any} data the DMP attempts to prefetch will be leaked.
Our adversary is assumed able to cause this DMP process to occur at any time, processing any memory location.

This model matches closely the behavior of the Apple M-series DMP which will dereference any value appearing in a cacheline that passes a heuristic check.
Prior work~\cite{gofetch,augury} building attacks using that DMP also demonstrate that a similar threat model is appropriate.

\begin{figure*}[h]
  \centering
  \includegraphics[width=.95\textwidth]{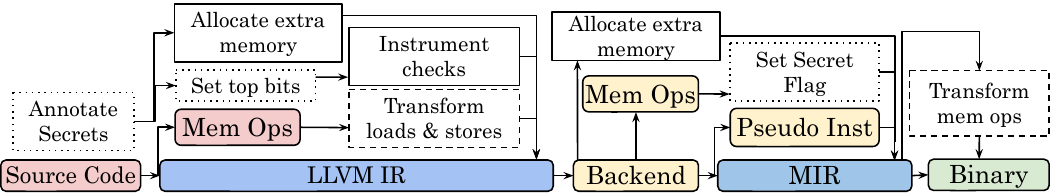}
  \caption{
The end-to-end pipeline of \toolname{}.
Extra memory is allocated for annotated secrets, with the top bits of addresses set for dynamic tracking.
At the LLVM IR level, memory operations are instrumented with checks and transformed loads and stores for our defense.
Memory operations generated by the backend are flagged or lowered to pseudo instructions that are later managed using an MIR pass.
}
  \label{fig:pipeline}
\end{figure*}

\section{Design}
\label{ssect:design}

The core of all DMP-enabled side-channels is allowing secret bits to flow into virtual address translation and (optionally) into an update to the cache state as part of an attempted prefetch.
Our defense rewrites memory stores and loads so that secrets are only present in the low half of 64-bit aligned memory regions, and they will never be used as part of a page walk due to the chosen high half prefix.

The specific prefix choice can be either an unmapped region of virtual memory, or an invalid address prefix that will never result in a prefetch.
For example, using a \code{0x00FF0000} prefix on ARM64 when using the standard 48-bit virtual addressing mode will result in no prefetch being attempted.

\begin{figure}[h]
  \centering
  \includegraphics[width=\columnwidth]{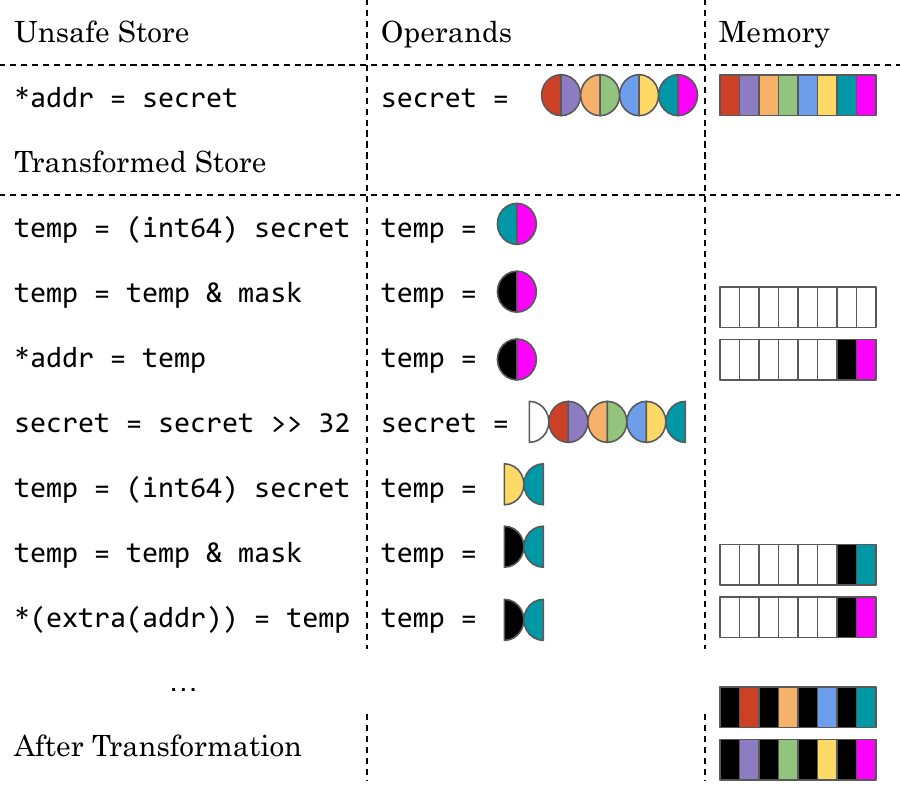}
  \caption{Example of transforming a 256-bit unsafe store. Each circle represents 64 bits. 64-bit chunks of the input (e.g the magenta and blue semi-circles) are split and padded with a 32-bit fixed constant value (black semi-circle.) The lower half with padding is stored at the original memory location and the upper half with padding is stored at extra memory allocated for the secret.}
  \label{fig:transform}
\end{figure}

\toolname{} is a compiler-based transformation defense against side channels introduced by \DMP{}s by ensuring no secret passes into a page walk.
For example, see the code in Figure~\ref{fig:ctswap-code} that shows a constant-time program that would require transformation for safety against DMP attacks.

To achieve this, we split and pad secret data into 32-bit segments, each stored separately along with a 32-bit fixed prefix which prevents it from resembling a valid address.
For 64-bit data, this method divides it into an upper and lower 32-bit value ($v_u$, $v_l$), which we then store in two 64-bit chunks with a fixed prefix $p$ as $p || v_u$ and $p || v_l$.
We adopt this approach because addresses are 64 bits in length and a fixed 32-bit prefix is sufficient to force failure of a page walk or even starting the prefetch.
We illustrate the transformation of the store instruction in Figure~\ref{fig:transform}.
Depending on the architecture, our technique can be tailored by determining the size of the prefix which will force failure of the validity check and avoid activation of the DMP or leaking data when a page walk fails.

We transform all stores that include secret data.
Our transformations are not functionally equivalent, so load operations must also be transformed to correctly retrieve stored data.
Since only data in memory needs protection against side channels induced by \DMP{}, registers and the CPU can hold secrets in unmodified form without risk of being leaked through data-at-rest attacks.
Therefore, we need not transform any non-memory operation using the data.

During execution time, some components of our defense require support for checking if a particular store is secret, allocation of additional memory for secrets, and keeping a track of that memory. We link this extra functionality into hardened binaries and refer to it as the \emph{runtime}.

\begin{figure}
\begin{minted}[fontsize=\small, breaklines]{c}
  void ctswap(uint64_t condition,
              uint64_t *a, uint64_t *b):
    uint64_t val_a = load(a)
    uint64_t val_b = load(b)
    uint64_t xor = val_a ^ val_b
    uint64_t mask = ~(condition-1)
    xor = xor & mask
    val_a = val_a ^ xor
    store(a, val_a)
    val_b = val_b ^ xor
    store(b, val_b)
\end{minted}
\caption{Example of a 64-bit constant-time conditional swap that would require transformation.
}~\label{fig:ctswap-code}
\end{figure}

\subsection{Loads and Stores}

Our operation is straightforward for values $\leq$ 32 bits: we allocate 64 bits of space, fill the upper half with our prefix, and put the stored value in the lower 32 bits.
A subsequent load to that address only needs to load the lower 32 bits to recover the true value.

\begin{figure}[h]

$Store(secret\_addr):$ 
\begin{minted}[fontsize=\small, breaklines]{c}
if (isSecret(secret_addr)):
  addr = unsetBit(secret_addr)
  for each 4-byte chunk in [addr, addr + size):
    chunk_addr = isShadow(addr) ?
                 heapAddr(addr) : addr
    *chunk_addr = merge(extractBits(value, chunk),
                        Prefix)
\end{minted}

$Load(secret\_addr):$
\begin{minted}[fontsize=\small, breaklines]{c}
if (isSecret(secret_addr)):
  addr = unsetBit(secret_addr)
  value = 0
  for each 4-byte chunk in [addr, addr + size):
    chunk_addr = isShadow(addr) ?
                 heapAddr(addr) : addr
    chunk = extractBits(*chunk_addr)
    value = insertBits(value, chunk)
  return value
\end{minted}

\caption{Pseudocode for transformed load and store operations of secret data $\geq 4$ bytes. These transformations assume that extra memory has been allocated for the secret data as needed. For both transforms, \texttt{isSecret()} checks if the address  has a high-bit set to mark it as secret, \texttt{unsetBit()} clears the secret tracking bit, \texttt{isShadow()} checks if the address points to extra allocated memory, \texttt{heapAddr()} translates a normally allocated address to the corresponding address in the extra memory. For \texttt{Store}, \texttt{extractBits()} extracts 32-bit chunks that will be merged with the special prefix using \texttt{merge()}. For \texttt{Load}, \texttt{extractBits()} extracts 32-bit chunks that were stored with the prefix, and \texttt{insertBits()} combines the extracted chunks into the final value. }~\label{fig:load-store-code}
\end{figure}

Handling stores of 64 bits and larger allocations is non-trivial, as it requires splitting stored value(s) into multiple discrete 64-bit prefixed components.
Depending on the secret labels, the loads and stores present will each need to be split into two operations.

For 64-bit stores, we split the stored value into two 32-bit chunks as shown in Figure~\ref{fig:load-store-code}.
To recover the value on load, we must load from both the original and additional allocations, then recombine the halves after removing the prefix as shown in Figure~\ref{fig:load-store-code}.

This becomes more complex for allocations that can be accessed element wise like arrays.
In such cases, we will need to allocate additional memory and overflow the values into this region if it is contiguous with the original allocation, or track an address for the non-contiguous region.
Each 4-byte chunk is then processed, storing alternately at the original and additional allocations along with the prefix.

Naively, this results in complex additional calculations for the new offsets into the array, regardless of where the additional allocation is located.
To avoid this complexity, we split the data element wise: the first 64 bits will be stored split across the first 64 bits of the original allocation and the additional allocation, simplifying offset computations.
This approach will also require keeping track of either the length or address and length of the additional allocation.
The only time a simple overflowing of data into an adjacent allocated memory region is feasible is when the size of the allocation is fixed and small like registers.

\section{Implementation}

\label{ssect:implementation}
In Figure~\ref{fig:pipeline}, we present the compilation pipeline, which takes the source program annotated with secrets and generates a binary that ensures no secret can leak from a DMP side-channel. We implemented it in LLVM as a transformation pass which runs on LLVM IR and handles memory operations generated from the source program. We also implement a MIR pass specifically for AArch64 backend which handles memory operations implicitly generated by the compiler for handling function calls, passing arguments by stack and spilling and filling registers. We also made changes to update the calling convention, spill sizes and support in the compiler at various places to acknowledge the updated spill size, reserved registers and changed calling conventions. Our LLVM transformation passes are written in 1850 lines of C++ and the runtime is written in 652 lines of C++.

\subsection{Source-level memory store handling}
The memory operations directly generated from the source code are transformed by \toolname{} at LLVM IR.
\subsubsection{Annotating and Tracking Secrets}
\label{tag-secret}

Transforming all memory operations under the assumption everything is a secret is not preferable, as these transformations incur overhead. Therefore, allowing for selective application of these transformations, \toolname{} supports annotating memory allocations as secret.

\noindent\textbf{Annotation Mechanisms.} Developers annotate secrets in source code using three mechanisms: (1) For \textit{heap allocations}, programmers replace standard \texttt{malloc/free} calls with \toolname{}'s custom allocator and deallocator wrappers (e.g., \texttt{secret\_malloc} and \texttt{secret\_free}). (2) For \textit{stack allocations}, developers add an attribute to variable declarations (e.g., \texttt{\_\_attribute\_\_((secret))}) to mark stack variables as secret. (3) For \textit{global variables}, \toolname{} provides a compiler flag to treat all globals as secrets, though individual global annotation is not supported on AArch64 due to Mach-O format limitations (discussed in Section~\ref{ssect:limit}). We transform memory operations that access locations annotated as secret during allocation in the source program.

Since these transformations alter operational semantics, we cannot modify a store operation without also handling all corresponding load operations that access the same address where the data was stored.
Achieving such precision is not feasible through static analysis alone, so we rely on dynamic secret tracking by using the upper bits of pointers as a tag.
For AArch64 architecture, only the lower 48 bits are used for addressing~\footnote{We do not support the uncommon 52/56-bit addressing modes.}, while the top byte is ignored, allowing us to use remaining 8 bits for tagging secrets.
This method hypothetically supports 256 types of alternate memory formats for data within the same binary.
In our implementation, we utilize a single conservative memory format where a secret is divided into 32-bit segments separated by 32 bits of fixed constant aligned such that every 64-bit chunk starts with the constant.
This ensures any 64-bit aligned location in memory containing a secret never appears as a valid address.
Our method allows multiple formats to co-exist, represented by different bit masks at the top bits of the address, but currently only uses a single bit to indicate secrecy.

Pointer authentication (PAC)~\cite{armpac} uses the same high address bits for signing pointers, as top bits are ignored during address translation and no other non-address bits are available.
In systems where PAC is enabled, our approach can integrate with it by signing all secret-related addresses with a specific PAC key.
This ensures compatibility between PAC and our defense mechanism.
However, we deliberately chose not to depend on PAC to avoid limiting our defense to systems where PAC is available or enabled.

\subsubsection{Allocating Memory for Secrets}

The \toolname{} transformation necessitates extra memory to store data along with constant, effectively doubling the memory required for secrets, except in cases involving data smaller than 32 bits. For heap allocations, this extra memory is provided by the custom allocator. For stack allocations, the additional memory is allocated on the heap by a runtime function. For global variables, we create a constructor-like function, called during initialization, which allocates additional memory in the heap and initializes it by zeroing to mimic the behavior of a variable stored globally.

\subsubsection{Setting Bits for Secrets}

To allocate a secret on the heap, the source program must use \toolname{}'s wrappers around memory allocators and deallocators. These wrappers allocate additional memory necessary for storing secrets along with the constant prefix in the heap, based on their sizes.
The extra memory allocation need not be next to the original allocation.
These memory wrappers are part of our runtime responsible for tracking the addresses of extra allocations and mapping them with the original allocation's address. During program execution, the runtime is queried to get the addresses of the extra allocation corresponding to the address of the data. The allocator wrappers also call the runtime to set appropriate bit in the address to mark them as secrets. If an untransformed store or load attempts to access an address returned by these allocators, it will crash due to the high-bit of the address being set, thus safely handling a failed transformation or unexpected third-party code interactions.

In LLVM, stack allocation is performed using the \code{alloca} instruction. To manage this, we instrument all functions to track each frame, mimicking the push and pop actions typical of function calls. For each function, we transform all alloca operations and add calls to the runtime, which takes the allocated address for the stack. The runtime then allocates extra memory required for storing the secret on the heap, handles tracking segments of secrets and their memory locations. It marks the allocation as secret by setting the upper bits of the address and replacing all uses of the address returned by alloca with this modified address. 

Since both \code{alloca} and \code{malloc} return a pointer, it is straightforward to set the high bits for tracking throughout the program. However, despite globals being pointers they are not returned by any function in LLVM IR, they simply exist. To set their high bits, we extend the logic for symbol relocation, which typically only supports small offsets for increments and decrements. Using operations like add with a sufficiently large offset can mimic setting the high bit with the appropriate offset. Unfortunately, this approach does not work for AArch64 due to limitations introduced by the Mach-O file format; further details are discussed in limitations, Section~\ref{ssect:limit}. Although we support secrets as globals, we do not allow individual globals to be annotated as secrets on AArch64. Instead, we provide the option to designate either all globals as secrets or none.

\subsubsection{Transforming Secrets}
\label{ssect:implementation:source_secret}

To transform secrets, we first add a runtime check for each memory operation's address to determine if it is marked as a secret by examining the bit pattern in the upper bits of the address. Since this verification occurs at runtime, we have to include code to handle both cases: when the data is a secret and when it is not. If a memory operation uses secret, we transform it within that specific code path.

The data is divided into 32-bit segments, with every alternate 32-bit segment stored in the extra memory allocated for that secret.
This organization simplifies tracking the data in the extra memory with minimal overhead, as we only need to know the starting address of an allocation in the extra memory corresponding to the secret allocation.
Our memory splitting method is supported for all data types, including vectors, arrays, characters, and integers, with the exception of floats and doubles. The exclusion of floats and doubles is due to operations like bit shifts that are not supported for floating-point arguments in LLVM IR and are required  during our transformation. Fortunately, floats and doubles are generally not used for storing secrets. We issue a compile-time warning if a memory operation is using a floating-point. We still transform them after casting them as integers. If a secret is not a floating-point it will never use this program path during execution.

Memory operations working with common sizes, such as 8, 16, 32, 64, 128, and 512 bits, are directly supported by the runtime. Calls to the runtime functions of appropriate sizes are added during transformation. We also handle unaligned accesses and non-standard sizes by directly transforming the LLVM IR. 

To ensure all memory operations related to secret allocations are transformed for correctness, we execute this transformation pass as the final step in the LLVM IR pass pipeline. This approach accounts for any stores or loads that might have been generated during optimization. Additionally, we disable certain passes, like the loop idiom pass, which could obscure memory operations from our transformation or address them in other parts of the compilation pipeline.

\subsubsection{Intercepting libc}
\label{ssect:impl:libc}
Our transformation modifies how data is stored and accessed in memory, requiring third-party code or libraries to be recompiled with the transformation to handle data correctly. One such library is libc, which must be transformed to ensure compatibility with our system. Libc contains two types of functions: those that could be implemented by a client (e.g., memcpy), and those that provide features that would otherwise be unimplementable as portable C code like syscall wrappers such as fopen. Syscalls can not work correctly with the pointers to modified data because we do not transform the kernel. Instead, we transform the data to the original format just before the syscall ensuring correctness and minimizing the exposure. Inline assembly is also heavily used in libc for performance optimization in functions like memcpy. It cannot be transformed by LLVM so to ensure compatibility we rewrite libc functions without inline assembly wherever possible. This reference version of commonly used libc functions that can be compiled using our transformation passes. Calls to libc functions are intercepted and redirected to this transformed reference implementation.

Any libc implementation without inline assembly and that can be compiled using Clang can be used in place of our reference implementation and transformed by our tooling.

\subsection{Compiler-Generated Memory Operations}

While source programs perform memory allocations and operations, these are not the only memory interactions a compiled binary might have. Compilers also perform memory operations, particularly on the stack, adding further memory transactions. While we are unaware of any \DMP{}-based attacks currently exploiting compiler-added operations to leak secrets, there is no reason they cannot be done. Thus, \toolname{} supports scenarios where secrets might be stored via memory operations introduced indirectly by the compiler. These operations include loads and stores for callee-saved registers, register spilling and filling, and argument passing through the stack.

\subsubsection{Annotating and Tracking Secrets}
\label{sssec:annotating}
With dynamic secret tracking, we can accurately determine during program execution whether a memory operation requires transformation.
Since compiler-transformed operations can't be fully known in advance and appear indistinguishable from source-generated operations, annotation must occur within the compiler but be conservative.
We transform all compiler-generated memory operations, except those related to the stack guard.
Stack guard values can instead be modified to choose a canary value that does not pass the DMP address checks, similar to our chosen prefix.

Compiler-generated memory operations are added during architecture-specific code generation at various stages of the pipeline.
LLVM's code generation begins with a directed acyclic graph, where memory operation nodes possibly containing secrets are tagged by adding a new metadata.
These nodes are transformed into Machine IR (MIR), losing metadata during the lowering process.
Directly transforming each operation would be complex and error-prone due to multiple transformation points.
Instead, we lower nodes with secret metadata to custom pseudo-instructions for secrets.
MIR stores and loads, directly generated as MIR instructions, are marked with a metadata flag.
We transform all memory operations identified as pseudo-instructions or with the secret flag at the end of the pipeline just before emitting assembly.

\subsubsection{Allocating Memory for Secrets}

We systematically expand the allocation size for each class of memory operations we aim to manage. For register spills and fills, we adjust the spill size for all AArch64 registers. Since the spill size becomes larger than the register size, this necessitates updating LLVM internal use of register and spill sizes. For stack-passed arguments, we modify the calling convention to use appropriate register sizes. Correspondingly, we ensure allocation sizes are adjusted, acknowledged by all dependent compiler components.

These allocations and transformations occur late in the code generation pipeline, often post-register allocation, complicating these transformations. We resolve this by placing extra allocated memory adjacent to the existing memory allocations, achieved through parameter adjustments responsible for original allocations or by manually updating allocation arguments. This approach seamlessly integrates expanded memory needs while maintaining alignment with the compiler's operations.

\subsubsection{Transforming Secrets}

While we specifically handle memory operations generated directly by the source code, applying similar transformations to those generated by the compiler during code generation is more complicated. Transforming these memory operations late in the code generation pipeline introduces constraints, such as difficulties in adding runtime calls and the necessity of using reserved registers as temporary storage during transformations. Given these constraints and the high number of potential transformation points, we adapt our approach to be lightweight and simple.

To streamline our implementation, we store adjacent memory segments directly without involving the runtime. We modify all the memory allocations inside the backend where secrets can be stored and transform each type of load and store instruction under the assumption that extra memory is contiguous. This means the offset calculation happens in the transformed code, rather than maintaining the same offsets as in the original and new allocations. We can afford this simplification because register sizes and the memory required for tasks like spilling, saving callee-saved registers, and passing arguments through the stack are well-defined and limited.

These backend transformations must be executed after all other backend passes, which could introduce additional memory operations. These transformations require scratch registers, so we reserve two 64-bit registers inside the compiler.

\subsection{Correctness, Security and Soundness}
Together, the properties below ensure that secrets are protected from DMP-enabled attacks without affecting the original behavior of the program.

\subsubsection{Correctness}
Correctness means that the externally visible behavior of the transformed program matches that of the original. For \toolname{}, that requires each transformed, deterministic, \emph{load} results in the same value as in the untransformed program.

To validate this, we rely on extensive testing with both real applications and our own test suite.
Our custom test suite includes targeted tests for specific features of our implementation, with over one hundred unit tests in total.
Additionally, we test our transformations on the cryptographic library libsodium~\cite{libsodium}.
We treat all memory allocations—heap, stack, and global as secret to maximize transformed stores and loads under diverse scenarios such as register spills, callee-saved registers, and stack-passed arguments.
Then, we use libsodium's own test suite to stress-test our transformations.

Stress-testing cannot guarantee correctness, but it increases confidence by exposing potential issues such as untransformed instructions or mismatched memory allocations.
For example, missing a transformation for a secret-related memory operation would likely lead to an incorrect memory access or program failure.
Our implementation passes all of our test suite, all of the libsodium test suite, and correctness tests in the microbenchmarks using libsodium for evaluation.

\subsubsection{Security}
Security means any modified store's value stored will never result in a successful page walk from the prefetcher.
Our transformations are secure by design, ensuring any transformed store results in memory containing one or more 64-bit-aligned chunks, each with a fixed 32-bit prefix. 

We specifically consider the Apple M-series DMP, running a MacOS-based system.
GoFetch~\cite{gofetch} (Section 4) comprehensively reverse-engineers the Apple M-series DMP, and identifies several heuristics it uses to identify candidate addresses for prefetch.
This implementation relies on the DMP's requirement that the storage location of a value must be in the same 4 GByte-aligned region as the interpretation of that value as a pointer.
So, a prefix that is guaranteed to result in an ``address'' that is 4GB away from any user allocation is sufficient as the DMP will only dereference pointers within 4GB of their storage location~\cite{gofetch}. We use a prefix of $0xdeadceef$ which satisfies this requirement as any ``address'' with this prefix is more than 4GB above the highest allowed virtual memory address in XNU~\footnote{\url{https://github.com/apple-oss-distributions/xnu/blob/8d741a5de7ff4191bf97d57b9f54c2f6d4a15585/osfmk/mach/arm/vm_param.h\#L137}}.
A similar selection approach would work for the Intel DMP, which also has a bounded prefetchable region relative to the storage location of the pointer.
Changing the specific prefix used is trivial in our implementation and could even be chosen at execution time.
An attacker might try to arrange allocations so that a prefixed 64-bit word appears to fall inside a DMP's valid locality window, or partially overwrite the high bits of a transformed chunk to synthesize a plausible pointer. Our chosen prefix places the synthetic address outside any user-mappable region the DMP will dereference (e.g., beyond the 4 GB locality range), so reshaping heap or stack layout cannot pull it into scope without violating virtual memory constraints. Forging a usable address would require corrupting the constant high 32 bits for multiple split chunks, which reduces to a memory safety attack against the application rather than a prefetch side channel. If future heuristics broaden validity, the defense can rotate prefixes or further shrink usable secret fragments, preserving the invariant that transformed words remain invalid prefetch candidates.

\subsubsection{Soundness} Soundness guarantees that all operations interacting with secret or secret-derived memory locations are consistently transformed.
For \toolname{}, we allow either manual annotations of secrets or complete transformation of all stored values if annotation is not feasible.
To validate that \toolname{} transforms all secret-marked memory regions, we performed a dynamic instrumentation experiment on macOS/AArch64 using QBDI~\cite{qbdi} focused on a hardened constant-time 64-bit conditional swap.
We instrumented every in-function memory write via a preload, classifying 8-byte stores into 16-byte clusters with the constant prefix values at offsets 0 and 8 (both \code{0xdeadceef}) and data-dependent LOW/HIGH components at offsets 4 and 12.
Across multiple runs we observed complete coverage with zero raw 64-bit, byte-sized, or otherwise uninstrumented stores providing evidence that secret-marked memory operations are consistently transformed.

\section{Evaluation}
\label{sect:eval}

\begin{table}[htbp]
\centering
\small  %
\begin{tabular}{>{\raggedright\arraybackslash}p{0.4\columnwidth}p{0.1\columnwidth}p{0.15\columnwidth}p{0.15\columnwidth}}
\toprule
Transformed memory operations in libsodium & libc used & Compile Time (s) & Binary Size (KB) \\
\midrule
None & glibc & 18.54 & 397.63 \\
None & libc$\star$ & 18.64 & 397.69 \\
Source & libc & 18.94 & 720.72 \\
Source$\scalebox{0.5}{\CIRCLE}$ & libc & 19.27 & 749.30 \\
Compiler $\And$ Source & libc & 18.98 & 1121.24 \\ 
Compiler $\And$ Source$\scalebox{0.5}{\CIRCLE}$& libc & 19.40 & 1193.85 \\
\bottomrule
\end{tabular}
\caption{Compile time and binary size analysis of the libsodium library with \toolname{}. Source and Compiler refer to the origin of memory operations, $\scalebox{0.5}{\CIRCLE}$ indicates all source allocations are secrets, libc is our reference glibc implementation, and $\star$ denotes untransformed libc.}
\label{compiler}
\end{table}

\begin{table*}[h!]
    \centering
    \renewcommand{\arraystretch}{1.2}  %
    \begin{tabular}{p{0.22\columnwidth}*{8}{p{0.070\columnwidth}}*{8}{p{0.060\columnwidth}}}
        \toprule
        \multirow{2}{*}{libsodium} & \multicolumn{8}{c}{Execution time} & \multicolumn{8}{c}{Memory Usage} \\  
        \cmidrule(lr){2-9} \cmidrule(lr){10-17}
        function 
        & B2 & B3 & S$\scalebox{0.5}{\Circle}$ & S$\scalebox{0.5}{\LEFTcircle}$ & S$\scalebox{0.5}{\CIRCLE}$ & SC$\scalebox{0.5}{\Circle}$ & SC$\scalebox{0.5}{\LEFTcircle}$ & SC$\scalebox{0.5}{\CIRCLE}$ 
        & B2 & B3 & S$\scalebox{0.5}{\Circle}$ & S$\scalebox{0.5}{\LEFTcircle}$ & S$\scalebox{0.5}{\CIRCLE}$ & SC$\scalebox{0.5}{\Circle}$ & SC$\scalebox{0.5}{\LEFTcircle}$ & SC$\scalebox{0.5}{\CIRCLE}$ \\
        \midrule
        xchacha20\textit{-e} 
                   & 1 & 1.03 & 1.38 & 1.48 & 12.38 & 1.44 & 1.48 & 12.29 
                   & 1.16 & 1.16 & 1.2 & 1.2 & 3.62 & 1.22 & 1.22 & 3.66 \\
        xchacha20\textit{-d} 
                 & 1 & 1 & 1.42 & 1.48 & 13.83 & 1.48 & 1.51 & 13.62 
                 & 1.16 & 1.16 & 1.2 & 1.2 & 3.62 & 1.22 & 1.22 & 3.66 \\
       xsalsa20\textit{-m}
                  & 1 & 1.03 & 3.54 & 3.53 & 41.64 & 3.59 & 3.59 & 41.02 
                  & 1.16 & 1.16 & 1.19 & 1.19 & 6.2 & 1.21 & 1.21 & 6.25 \\
       xsalsa20\textit{-v}
                  & 1 & 1.03 & 3.54 & 3.6 & 43.36 & 3.63 & 3.64 & 42.5 
                  & 1.16 & 1.16 & 1.19 & 1.19 & 6.2 & 1.21 & 1.21 & 6.25 \\
       salsa20\textit{-e}
                  & 1 & 1 & 1.64 & 1.64 & 18.64 & 1.64 & 1.64 & 18.61 
                  & 1.16 & 1.16 & 1.19 & 1.19 & 1.72 & 1.21 & 1.21 & 1.77 \\
       salsa20\textit{-d} 
                  & 1 & 1 & 1.64 & 1.64 & 18.82 & 1.64 & 1.64 & 18.77
                  & 1.16 & 1.16 & 1.19 & 1.19 & 1.72 & 1.21 & 1.21 & 1.77 \\
       blake2b\textit{-h}
                  & 0.98 & 1.24 & 1.24 & 1.24 & 1.24 & 1.32 & 1.32 & 1.32
                  & 1 & 1 & 1 & 1 & 1 & 0.98 & 0.98 & 0.98 \\
       \textit{h-}sha256\textit{-m}
                  & 0.99 & 1.02 & 3.50 & 3.49 & 41.21 & 3.55 & 3.55 & 40.66
                  & 1.16 & 1.16 & 1.19 & 1.19 & 6.2 & 1.21 & 1.21 & 6.25 \\
       \textit{h-}sha256\textit{-v}
                  & 0.99 & 1.02 & 3.57 & 3.56 & 42.82 & 3.60 & 3.60 & 42.09
                  & 1.16 & 1.16 & 1.19 & 1.19 & 6.2 & 1.21 & 1.21 & 6.25 \\
       \textit{h-}sha512\textit{-m}
                  & 1 & 1.10 & 3.82 & 3.94 & 69.71 & 3.44 & 3.53 & 69.15
                  & 1.16 & 1.16 & 1.19 & 1.19 & 10.55 & 1.2 & 1.2 & 10.6 \\ 
       \textit{h-}sha512-\textit{v}
                  & 1 & 1.08 & 3.88 & 3.99 & 70.39 & 3.46 & 3.54 & 69.54
                  & 1.16 & 1.16 & 1.19 & 1.19 & 10.55 & 1.2 & 1.2 & 10.6 \\
       ed25519\textit{-s}
                  & 1 & 1.02 & 3.67 & 4.30 & 98.22 & 5.85 & 5.85 & 90.63
                  & 1.17 & 1.17 & 1.25 & 1.25 & 15.71 & 1.37 & 1.37 & 27.1 \\
       ed25519\textit{-v} 
                  & 1.01 & 1.01 & 2.43 & 2.44 & 67.83 & 3.36 & 3.37 & 68.46
                  & 1.17 & 1.17 & 1.25 & 1.25 & 15.71 & 1.37 & 1.37 & 27.1 \\
       aead\textit{-e}
                  & 1.02 & 1.04 & 1.48 & 1.48 & 13.10 & 1.52 & 1.54 & 13.02
                  & 1.17 & 1.17 & 1.2 & 1.2 & 3.37 & 1.23 & 1.23 & 3.41\\
       aead\textit{-d}
                  & 1 & 1.02 & 1.51 & 1.51 & 14.25 & 1.55 & 1.55 & 14.15
                  & 1.17 & 1.17 & 1.2 & 1.2 & 3.37 & 1.23 & 1.23 & 3.41\\
        \bottomrule
    \end{tabular}
    \caption{We measure the runtime overhead caused by our transformations. Results are multiplicative overhead relative to a baseline LLVM build of libsodium with none of our passes or changes. \textit{B2} and \textit{B3} are the versions of the baseline which uses the reference implementation of libc not transformed and transformed with our transformations respectively. \textit{S} represents the overhead when only the secrets from the source are handled and SC when the operations generated from the compiler are also handled. We use $\scalebox{0.9}{\Circle}$, $\scalebox{0.9}{\LEFTcircle}$ and $\scalebox{0.9}{\CIRCLE}$ to show that the source code is annotated with no secret, selectively annotated with secrets, everything is assumed to be a secret respectively. The prefix \textit{h-} represents hmac and the suffix \textit{-e}, \textit{-d}, \textit{-m}, \textit{-v} and \textit{-s} represents \textit{encrypt}, \textit{decrypt}, \textit{mac}, \textit{verify} and \textit{sign} respectively. We also note that the aead used in experiments is chacha20 poly1305.}
    \label{tab:performance_memory_usage}
\end{table*}

\toolname{} affects aspects of the target program and the compiler. In this section, we evaluate four dimensions of the overheads introduced by our defense:

\begin{itemize}
    \item Compile-time (Section~\ref{ssect:compilation-overhead})
    \item Binary size (Section~\ref{ssect:binary-size-impact})
    \item Runtime performance implications (Section~\ref{ssect:runtime-overhead})
    \item Runtime memory consumption (Section~\ref{ssect:memory-consumption})
\end{itemize}

Our evaluation is conducted using the libsodium library, a comprehensive cryptographic library offering a wide array of cryptographic primitives and APIs. One limitation of using LLVM for our implementation is that we cannot modify inline assembly. Thus, we ensure \texttt{--disable-asm} (used by WebAssembly targets) to force reference implementations for most operations. We also disable the use of variable length arrays (see Limitations Section \ref{ssect:limit:dyn_stack}). Other than this, we build libsodium using its usual compilation flags. We ensure libsodium's test suite passes all tests for every build.

Our defense necessitates recompilation of libraries the program relies on, integrating our transformation to ensure library code manages annotated secrets effectively and correctly handles memory operations pertaining to these secrets. Given that popular libc implementations often include inline assembly for optimizations and may lack options for disabling them, as seen in libsodium, we developed a reference implementation of libc without inline assembly. This implementation was then compiled using our transformation passes such that data is correctly passed to system calls.

Since most applications do not spend the majority of execution time performing cryptographic operations, we analyze the overheads of libsodium using our microbenchmarks. All evaluation (compilation and execution) is performed on a stock Apple Mac Mini with an Apple M1 chip (8-core CPU), 8 GB of unified memory, and storage on SSDs.

\subsection{Compile Time Overhead}
\label{ssect:compilation-overhead}

\toolname{} includes passes to process memory operations at both the LLVM IR and MIR levels, which can increase compilation time. Table \ref{compiler} illustrates wall-clock time for compiling libsodium using a single job, specified with \texttt{-j 1}. Our baseline comparison is the LLVM compiler without modifications, such as reserved registers, disabled transformation passes for \toolname{}, and no interception of libc functions. We did clean builds of libsodium 10 times and report average compilation time.

To understand impact of our reference implementation of libc on the baseline, we compiled libsodium with a transformation for intercepting libc functions. It does not matter whether libc is built with our transformations or not, as the compiler only needs to add calls to it. We observed a 0.53\% increase in performance overhead due to interception of glibc functions. The transformed libsodium cannot work with an untransformed libc, as explained in Section~\ref{ssect:impl:libc}. Compile time increased by 2.15\% when all source-generated secrets are transformed, and by 2.37\% when compiler-generated memory operations are also handled. This increase is attributed to time spent transforming memory operations and adding runtime calls for tracking secrets and their additional allocations. Next, we compiled libsodium assuming all allocations inside it are secret. We found an increase of 3.93\% and 4.63\% when only source-generated operations are modified and when both source- and compiler-generated operations are included, respectively. This is due to the increased number of runtime calls to track secrets and their allocations.

\subsection{Binary Size Impact}
\label{ssect:binary-size-impact}

The LLVM IR transformation passes in \toolname{} introduce runtime checks to determine if memory operations are handling sensitive data. If they are, a control flow path is established, invoking either runtime functions or a sequence of instructions. Furthermore, the MIR pass modifies flagged memory operations by splitting secrets during storage and recombining them upon loading, which increases binary size.

Table~\ref{compiler} presents binary size for the same baseline LLVM version used for understanding compile time overheads, as discussed in Section~\ref{ssect:compilation-overhead}. To understand impact of our reference implementation of libc on the baseline, we compiled libsodium with a transformation for intercepting libc functions. We found an increase of 0.01\% because some glibc calls are directly replaced with target-specific instructions by the compiler. We observed an increase of 81.25\% in binary size when our transformations for modifying secrets from source were enabled, as this adds checks and calls to the runtime function for every memory operation. Supporting compiler-generated memory operations also increases binary size by 181.98\% because most operations generated by the compiler are transformed into a series of instructions which are not further optimized. Additionally, we observed an increase in binary size by 88.44\% and 200.24\% respectively, when all allocations in libsodium are considered secret, due to more runtime calls required to track secrets and their allocations.

\subsection{Runtime Performance Overhead}
\label{ssect:runtime-overhead}

To evaluate the runtime performance overhead of \toolname{}, we developed eight microbenchmarks using the libsodium library. These benchmarks cover common cryptographic use cases, including stream and block ciphers, hash and key generation, and encryption algorithms, with source code annotated to indicate secrets. We compared performance of each cryptographic primitive across scenarios: running with the baseline compiler with standard libc and the reference libc implementation (as detailed in section \ref{ssect:runtime-overhead}); with the compiler configured to apply transformations to handle memory operations generated directly from the source code; and with transformations applied to both source and compiler-generated memory operations. We also compared source code with and without annotations (assuming there is no secret) and assuming everything is a secret to better understand overhead.

For each cryptographic primitive, we conducted an evaluation loop of 1,000 iterations, preceded by 25 warm-up iterations. Run times were measured using the AArch64 assembly instruction \texttt{msr} to access the virtual count register. The reported overheads were normalized against mean run time of libsodium compiled with baseline LLVM, with results presented in Table \ref{tab:performance_memory_usage}.

Our findings indicate an average 262\% slowdown when handling memory operations generated directly from the source code, and a 274\% slowdown when considering memory operations generated by the compiler.

\subsection{Memory Consumption Overhead}
\label{ssect:memory-consumption}

We utilize the same microbenchmarks previously employed to evaluate runtime costs, as detailed in Section \ref{ssect:runtime-overhead}. Our assessment focuses on comparing the maximum resident set size observed after 100 warm-up runs and 1000 evaluation runs, which reflects the peak physical memory occupied by a process during execution. We begin by examining impact of using the reference libc implementation, both with and without transformation, against the unintercepted libc paired with the baseline compiler, as outlined in Section \ref{ssect:compilation-overhead}. Our findings indicate a negligible increase in memory consumption attributed to the reference libc. This increase is due to simplistic algorithms used in various function implementations, which expanded significantly upon application of transformation to the reference libc. The change in memory consumption is illustrated in Table \ref{tab:performance_memory_usage}.

Furthermore, when enabling transformation for handling memory operations generated directly from the source code, as compared to the baseline compiler with the reference libc implementation, we observed a memory consumption increase of 118\%. Notably, when we assumed the program contained no secrets, the change in memory consumption was negligible. This observation underscores that memory required for storing alternative 32-bit segments of secrets and their tracking does not significantly impact overall memory consumption, given the sparse number of secrets.

Additionally, our analysis reveals the primary cause of increased memory consumption is the rise in runtime calls, leading to more register spilling. Equally contributing is the necessity to convert each data item back to its original form before passing it to a system call. This conversion involves allocating memory for data in its original format and using an instrumented memcpy to copy data from its transformed form. Moreover, enabling backend transformation for handling compiler-generated memory operations further increases memory consumption by 121\%. This is due to the requirement for double the memory allocation for each register that is spilled, saved by the callee, or passed as an argument.

\section{Discussion}

\toolname{} introduces noticeable overhead in compile time, binary size, execution time, and memory consumption for the program it protects. These results are expected, as we transform each memory operation into either a sequence of instructions or runtime calls. Even if the data being operated on is not secret, all compiler-generated memory operations except for obvious cases, such as those generated for stack guards are transformed, and a check for secret data is instrumented for all memory operations generated directly from the source code.

\subsection{Data Split Size}
The more specialized the splitting strategy is, based on architecture and microarchitecture, the less overhead it will incur.
In \toolname{}, we split all data into alternating segments of 32 bits and mask the top 32 bits with a fixed constant value, ensuring the data never resembles an address. For a 128-bit store, we now use 256-bit memory, along with four 64-bit instructions that take the 32-bit segments from the original value and mask the top half with fixed constant. This could have been reduced to 192-bit memory and three 64-bit stores if, instead of masking the top 32 bits with fixed constant, we masked the top 16 bits, thus reducing runtime and memory consumption overhead.

\subsection{Selective Transformations for Secrets}
Our goal is to selectively protect secrets from being leaked through data-dependent prefetcher-induced side channels. To achieve this, we instrument all functions to handle secrets, including common libc functions. However, this is not always necessary. For example, if a function only receives arguments through registers and these can be proven to never be secrets, those functions do not need transformation. Furthermore, if a function takes a pointer as an argument, and it can be shown that it does not point to secret-bearing memory, this function also does not need instrumentation. We leave these improvements to future work.

\subsubsection{Transforming Compiler-Generated Memory Ops}
The memory operations generated by the compiler for handling register spills, callee-saved registers, and arguments passed through the stack do not require handling, similar to those generated for stack guards, when it can be ensured the compiler will never store secrets in the stack. Techniques such as secure register allocation and inlining can be employed to avoid transforming compiler-generated memory operations.

Performing an analysis of memory operations to check their operation on secret data is not very useful after register allocation, unless it is known that a particular memory address range does not interact with secrets. In such cases, we would not need to allocate extra memory for the data.

\subsection{Declassification Boundary}

Our defense mechanism involves altering the way data is stored in memory so that data-dependent prefetchers cannot interpret them as addresses. When data is stored in memory, it is converted into a different form. However, it must eventually be transformed back, or declassified, when it is loaded from memory into registers. Once the data is in registers, those operating on the registers need not be concerned with the format in which it was originally stored in memory. This concept also applies to library functions, which need to understand how to process data in its transformed or classified state. Hence, libraries must be recompiled with instrumentation to appropriately handle data classified within memory.

This process creates a declassification boundary where data is reverted to its original form. Certain libraries and ABIs cannot be transformed to accommodate this instrumentation and, therefore, cannot manage classified data. We define registers as natural declassification boundaries for our mitigation strategy, as prefetchers operate based on data stored in memory—not data held in registers, which need not be classified. Similarly, we also treat system calls as declassification boundaries, as it is not feasible to instrument their implementation to manage classified data.

\section{Limitations}
\label{ssect:limit}

\subsection{Inline Assembly Support}
One significant limitation of our approach comes from the lack of compiler support for inline assembly. LLVM treats inline assembly and assembly files as strings with definitions and uses, but lacks support for transformations. This restricts us to defending only the reference implementations in libsodium, which do not use inline assembly. Often, such limitations can be addressed by transferring compiler transformations to a binary rewriter. However, this method falls short as memory allocated for secrets must be adjusted to accommodate changes in spill size and other compiler internal parameters, such as the size of callee-saved registers. A binary rewriter and analyzer are incapable of addressing these requirements. Future work could extend \toolname{} to provide suggested assembly outputs that an expert developer could further refine for performance optimization.

\subsection{Architecture Agnostic Implementation}
Our instrumentation pass, which manages memory operations directly generated from the source, operates on LLVM IR, making it architecture-agnostic. However, to handle memory operations added by the compiler, we needed an architecture-specific pass that works on MIR, specifically for the AArch64 backend. To support other backends, this implementation will need adaptation. While similar changes might be required, they will not be identical. For instance, in AArch64, the Mach-O compact unwind format mandates callee-saved registers be stored in adjacent register pairs.

\subsection{Precise Tracking of Secrets}
For accurate tracking of secrets throughout the program, it is essential to instrument only memory operations that interact with secrets. However, achieving completely precise alias analysis is not feasible due to limitations of static analysis. 
As any incorrectly classified load will break behavior, we opt for a dynamic tagging the address during allocation as discussed in Section~\ref{tag-secret}.

\subsubsection{Tracking of secret-derived values: }

Currently, secret-derived values must be handled and annotated similarly to cryptographic secrets, or all values treated as secrets. 
This means that for a function like memcpy both the source and destination needs to be annotated as secret if we want the secret data to be protected.
A better middle ground would include taint analysis for annotated secrets to identify derived values and mark them as secret as well.
This would also require adding an explicit declassify operation once a cryptographic operation is complete.
We leave these improvements to future work, and note overheads for ``consider all values secret'' are an upper bound on performance impact of our transformations.

\subsection{Annotation Support for Global Secrets}

Some features of \toolname{} are architecture-specific. In our prototype implementation for AArch64, annotation of global variables as secrets is not currently supported. However, this limitation is not fundamental and can be addressed with changes to the relocation code. For example, ARM's pointer authentication (PAC) already uses similar mechanisms to set higher address bits. We leave this support to future work as secrets are rarely stored in global variables.

For the x86 architecture using the ELF format, annotating global variables is supported. The symbol relocation logic can handle 64-bit offsets, allowing us to set higher bits of a global address and tag it as secret. On AArch64 with the Mach-O format, however, symbol relocation does not permit offsets that could be used to set higher bits. As a result, globals on AArch64 can only be considered secrets in their entirety or not at all. If marked as secret, they are fully hardened by our system. While this limitation applies specifically to tagging individual globals as secrets on AArch64, it does not affect overall functionality of \toolname{}. Future work could explore extending relocation support for AArch64 if needed, but given the rarity of storing secrets as global variables, we consider this a low-priority enhancement.

\subsection{Dynamic Stack Allocation}
\label{ssect:limit:dyn_stack}
Our prototype implementation does not support hardening variable-length stack arrays or the arguments to variable argument functions. These features require separate stack allocation within the compiler, necessitating additional support and careful bookkeeping to manage stack allocation. The allocated memory must also be tagged to track secrets. Fortunately, libsodium provides an option to disable variable-length arrays, which we have set for our builds. As for variable argument functions, these are only present in correctness tests of libsodium, where (for testing) we replaced \code{printf} calls with \code{puts}, thus avoiding use of variable arguments. We do support \code{alloca} which can be used to dynamically allocate memory on the stack if needed.

\section{Related Work}
In this section, we provide an overview of related
works in the context of this paper, which can be broadly categorized into three main areas: use of compilers for defending against side-channel attacks, existing techniques for modifying the data representation and software prefetcher attacks.

\subsection{Side-channel mitigations in compilers}
Compiler-based approaches are a common method for defending against side-channel attacks.
These techniques offer protection without hardware modifications, enhancing deployability across existing systems.

Concurrent work, \textit{Zebrafix}~\cite{zebrafix} proposed a compiler-based approach that interleaves secret memory with non-secret values to mitigate data-at-rest attacks.
In their case they use incrementing counters---rather than static prefixes---to mitigate silent store and ciphertext side-channels.
The authors suggest this could extend to cover DMPs via block size reduction, though implementation and evaluation only cover non-DMP attacks.

A common strategy is to enforce constant-time (CT) execution to eliminate timing side channels. This involves using analysis to find program points which can be transformed to make them independent of secrets~\cite{coppens2009practical, wu2018eliminating}. However, CT is not always sufficient to protect against side-channel attacks; sometimes leaking noise or wrong data is easier than protecting against the attack. Speculative Load Hardening (SLH), the defense for Spectre v1 implemented in LLVM, poisons the value during speculative execution~\cite{llvmSpeculativeLoad}. Rane et al.~\cite{rane2015raccoon} developed Raccoon, a compiler-based system that transforms programs to obfuscate memory access patterns, making them resistant to cache timing attacks. Sometimes the attack depends on a hardware feature, such as a particular arithmetic operation optimization, which can introduce a side-channel. These attacks can be foiled by either avoiding generating patterns that can trigger them or by transforming triggering patterns into safe ones~\cite{cio}.

Our approach in the compiler to defend against DMP is by avoiding triggering it, not through code patterns but by modifying data representation in memory.

\subsection{Similarity to Attacks on Stored Data}

While they use fundamentally different techniques to recover data, there are interested parallels to attacks that recover data stored in disks or RAM via external measurement.
A classic example would be cold boot attacks~\cite{lest_rem} which can similarly recover data presumed safe when not actively processed.
This suggests that strategies designed to protect stored data on disks or RAM could offer insights for protecting against DMP-enabled attacks.
Both contexts require securing static data against indirect inference attacks that bypass traditional access controls~\cite{flush_reload, uarch-side-channel}.

\subsubsection{Data masking} 
Data masking refers to representing data as a combination of multiple elements that cannot be correlated with the original data~\cite{goubin2011protecting}. When data is divided into $d$ pieces, it is called $dth$ order masking, where data is represented by $d$ shares~\cite{rivain2010provably}. Combining this masked data back to its original form can be achieved using binary representation techniques, often referred to as boolean masking~\cite{coron2013higher}. For example, xor-ing data with a constant is a common boolean masking technique~\cite{rivain2010provably}. Additionally, these masking techniques can also operate on decimal representations, such as Shamir's secret sharing~\cite{goubin2011protecting}.

Existing masking techniques cannot be used to defend against side channels induced by the DMP because they do not modify data to specifically fail address translation and will thus still leak partially. Unlike masking techniques, our defense restricts data modification to memory rather than registers, making it possible to efficiently implement in the compiler. This is one reason compiler-based higher-order data masking implementations are uncommon. Data masking techniques are significantly more common in hardware, often for protecting against power analysis side-channels.

\subsubsection{Data Space Randomization}

Data Space Randomization (DSR)~\cite{bhatkar2008data} randomizes data values in memory to prevent them from being predictable or resembling specific patterns. While DSR shares the conceptual goal of preventing data from resembling valid addresses, it has fundamental limitations for DMP defense. DSR applies probabilistic transformations (typically XOR with random masks), leaving a non-zero probability that randomized data could still resemble a valid address and trigger the DMP.
In contrast, \toolname{} provides deterministic guarantees by using architecturally-aware prefixes that provably fail address translation for any 64-bit aligned value. Moreover, DSR would likely incur higher overhead due to generating and managing random masks for each data item, whereas our fixed prefix approach requires only simple bit manipulation with a compile-time constant.

\subsection{Software Prefetching Attacks}
In a parallel line of work, multiple projects discovered that \emph{software} prefetching instructions can be abused by attackers as well.
For example \texttt{PREFETCHW} can be used for high-capacity covert channels and to enhance side-channel attacks through timing analysis~\cite{prefetcher_attacks_adversarial_prefetch, prefetcher_attacks_bypassing_smap}. Additionally, prefetchers can manipulate auxiliary components like branch predictors, allowing targeted cache evictions~\cite{prefetcher_attacks_bunnyhop}. Lastly, different architectures exhibit vulnerabilities; for example, Lipp et al.~\cite{prefetcher_attacks_amd} discuss timing leaks in AMD CPUs that affect kernel isolation, highlighting the broader challenge of securing all forms of prefetch mechanism.

\section{Conclusion}
In this paper, we introduced \toolname{}, a compiler-based framework designed to address the emerging threat of Data Memory-Dependent Prefetcher (DMP) attacks. Recognizing the limitations of constant-time programming against these sophisticated vulnerabilities, \toolname{} offers a novel approach by transforming memory operations to prevent secrets from being exposed as memory addresses. This effectively shields them from potential exploitation by prefetchers.

Drawing on lessons from previous microarchitectural threats like timing and Spectre attacks, our work highlights the importance of proactive compiler and software defenses to keep pace with advancing vulnerabilities. Implemented using LLVM, our solution demonstrates compatibility with both source-level and compiler-generated operations on the Apple M1's AArch64 architecture. Though our approach involves performance trade-offs, our analysis confirms that robust protection is achievable.

\toolname{} lays the groundwork for defending against future microarchitectural attacks, reinforcing the security of current systems. As microarchitectural complexity continues to grow, our framework provides a crucial defense layer, and we encourage further development of transformation strategies to mitigate potential side channels.

\bibliographystyle{plain}
\bibliography{bibs/references}

\end{document}